\documentclass[11pt]{article}

 \usepackage{natbib}
 \usepackage{setspace}
\usepackage[a4paper, total={6.5in, 9in}]{geometry}
\usepackage{url}

   \usepackage{graphicx}
   \usepackage{amsmath}
   \usepackage{footmisc}
   
   \makeatletter
   \def\@xfootnote[#1]{%
  \protected@xdef\@thefnmark{#1}%
  \@footnotemark\@footnotetext}
  \makeatother
\begin{document}

\title{The 2010 and 2014 floods in India and Pakistan: dynamical influences on vertical motion and precipitation}
\date{}

\maketitle

%% ------------------------------------------------------------------------ %%
%
%  TITLE
%
%% ------------------------------------------------------------------------ %%

%
% e.g., \title{Terrestrial ring current:
% Origin, formation, and decay $\alpha\beta\Gamma\Delta$}
%

%% ------------------------------------------------------------------------ %%
%
%  AUTHORS AND AFFILIATIONS
%
%% ------------------------------------------------------------------------ %%

%Use \author{\altaffilmark{}} and \altaffiltext{}

% \altaffilmark will produce footnote;
% matching \altaffiltext will appear at bottom of page.

\author{Daniel A. Shaevitz$^1$\footnote[*]{Email: das127@columbia.edu}, Ji Nie$^2$, and Adam H. Sobel$^{1,2}$
\\ \\
$^1$ Department of Applied Physics and Applied Mathematics, Columbia University, New York, NY, USA. \\
$^2$ Lamont Doherty Earth Observatory, Columbia University, New York, NY, USA.
}

\section*{Abstract}
Devastating 
floods in northeast Pakistan and northern India occurred in July 2010 and September 2014 as a consequence
of extreme precipitation events. The 2010 and 2014 flood events had similar synoptic flow patterns that led to an anomalously high moisture content in the flood region.  
The quasi-geostrophic omega equation is inverted in order to attribute components of the large-scale vertical motion profile
to synoptic forcing, diabatic heating, and mechanically forced orographic ascent. The results show that diabatic
heating is the dominant contributor to the large-scale vertical motion, and 
suggest that the orographic forcing by flow over the Himalayas is the dominant mechanism by
which the convection that produces the heating is forced. Analysis of a longer data record shows that instances of extreme
precipitation in this region over the last eleven years are closely associated with simultaneously
large values of orographic forcing and column-total precipitable water vapor.

%% ------------------------------------------------------------------------ %%
%
%  BEGIN ARTICLE
%
%% ------------------------------------------------------------------------ %%

% The body of the article must start with a \begin{article} command
%
% \end{article} must follow the references section, before the figures
%  and tables.

%% ------------------------------------------------------------------------ %%
%
%  TEXT
%
%% ------------------------------------------------------------------------ %%

\section{Introduction}
In early September 2014, near the end of the monsoon season, heavy rains caused landslides and flooding in the Jammu and Kashmir region of India and nearby regions of Pakistan, killing hundreds of people \citep{Mishra15}.  

\begin{figure}
%\centering
\includegraphics[width=25pc,angle=0]{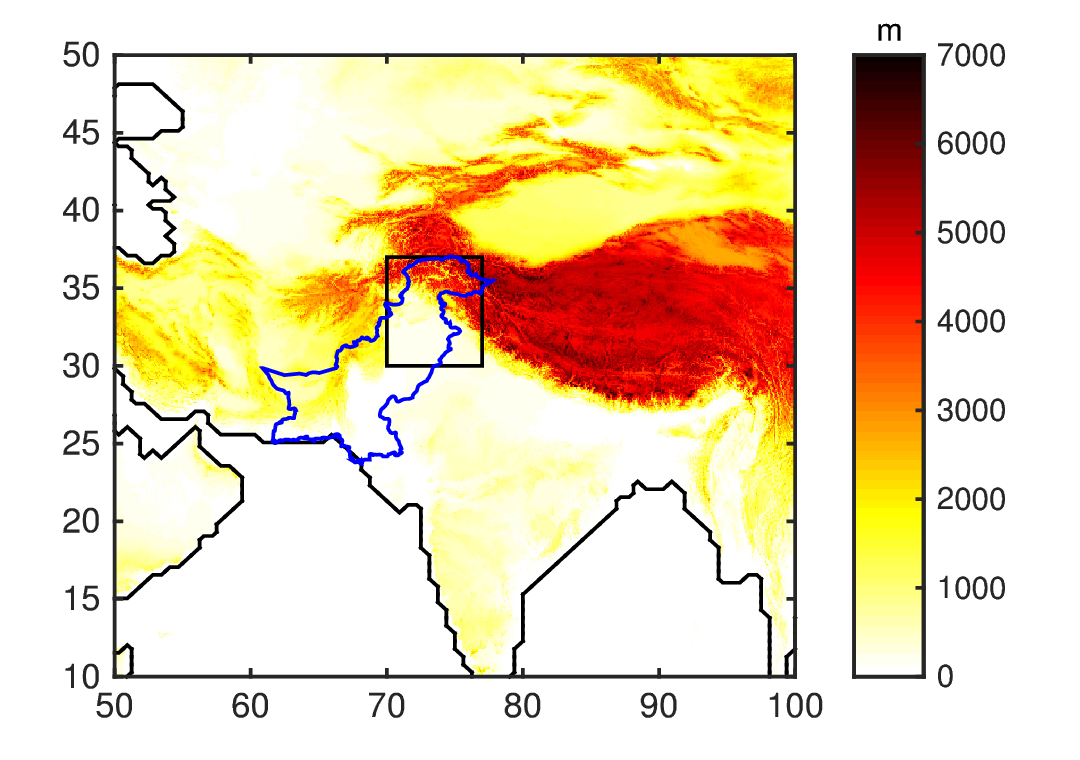}\\ 
\caption{Topography map.  The black rectangle represents the flood region and the blue outlines the border of Pakistan.}
  \label{topoMap}
\end{figure}

Extreme rainfall in roughly the same region (the precipitation region can be seen in figure \ref{topoMap}) led to historic floods of the Indus river basin in northeast Pakistan, submerging a significant fraction of the country and causing  over 2000 deaths \citep{Akthar11}.  The July 2010 floods were related to an extratropical blocking anticyclone event that also led to an intense heat wave in Russia \citep{Lau11}.  
A southeasterly flow was present over India during the end of July 2010, caused by an intense pressure gradient between an anticyclone over the Tibetan Plateau and a low pressure disturbance traveling westward across India \citep{Houze11,Rasmussen15}. This southeasterly flow brought moisture into northwest India and eastern Pakistan, where the moist air stream was lifted as it reached the slopes of the Himalaya.

We are interested in gaining a deeper understanding of the dynamics of these events of 2010 and 2014, as well as other similar events ({\it e.g.} \cite{Rahmatullah52} described a flood in the region during August 1949 that was also related to a southeasterly flow over India due to a mid-tropospheric trough).  Two necessary conditions for an extreme precipitation event are upward motion and a sufficient supply of moisture.  The events occurred in the western foothills of the Himalayas, a region with a very steep gradient of topography (figure \ref{topoMap}).  Steep topography gradients are well known to be able to produce orographic precipitation when moist air flow impinges on them \citep{Roe05,Smith06}. 
These events occur in a sufficiently warm, humid environment that the precipitation may be convective, at least in part, with some of the vertical motion connected to diabatic heating which allows air parcels to ascend across potential temperature surfaces. At the same time, upper-level disturbances can cause ascent through quasi-adiabatic potential vorticity dynamics. We attempt here to quantify the roles of each of these processes in the 2010 and 2014 events.

The events of 2010 were extensively studied in \cite{Martius13} and \cite{Galarneau12}.  It was concluded that the transport and convergence of moist air into the flood region was driven by monsoonal low-level flow features and that positive potential vorticity anomalies at upper levels, reaching Pakistan from the extratropics, induced a surface wind field that had a significant component directed orthogonal to the topographic barrier.  It was also suggested that the upward motion was mainly driven by the forcing of topography and that forced quasi-geostrophic ascent from the upper-level large-scale flow was fairly weak. In this study, we extend these two studies in three ways. First, we perform a more in-depth analysis of the factors influencing large-scale vertical motion in greater detail for the 2010 event, including examining the vertical profiles of vertical motion attributable to different factors via quasi-geostrophic omega equation. Second, we apply the same approach to the 2014 event.  Third, we place both events in longer historical context by examining the association of extreme precipitation events over a ten-year period with the factors identified as likely causal influences in the 2010 and 2014 events. 

The quasi-geostrophic omega equation \citep{Clough96,Gray06} is a useful framework for understanding and quantifying the factors associated with vertical motion outside
the deep tropics.
While the assumptions of quasi-geostrophic theory are not technically valid in this application, due to the large topographic variation in the region of interest as well as other possible violations, it is nonetheless a useful exercise to apply the theory and see how well it works.  The results in section \ref{qgOmega} demonstrate that the omega equation captures the dynamics quite well.  A great advantage of the quasi-geostrophic omega equation is that it is linear, allowing for direct decomposition of the vertical motion into components forced by the large-scale quasi-adiabatic dynamics, diabatic heating, and orography. A more exact nonlinear model would not  allow such a straightforward decomposition.  The inclusion of the contributions of diabatic heating and orography, not considered in previous work, allows for a direct comparison with the actual vertical velocity and thus provides confirmation of the validity of using this model for these events, as well as a more complete description of the vertical motion.

These precipitation events involve deep convection, large diabatic heating in the troposphere, 
and thus vertical motion directly associated with that heating in the omega equation. 
What is not so easily apparent is the role of each of the possible factors in causing the convection.  One possibility is that the large scale circulation induces upward motion that forces the convection.  Another possibility is that when surface winds are aligned towards the topographic barrier, the upward motion forced mechanically near the surface then triggers the convection.  To fully understand what is driving the vertical motion, it is necessary to understand the interactions of the large-scale circulation and topographically forced lifting with the convective heating. The current study is limited to analyzing observational data, takes the heating rate as a given, and thus cannot explicitly determine what
causes that heating. However, the analysis developed here is used to provide forcing terms which are used in the modeling study by \cite{Nie16b}, using the column quasi-geostrophic method of \cite{Nie16} in order to more directly separate the 
influences of  causal understanding of the controls exerted by orographic lifting and upper-level disturbances on the convection. Here we establish, as necessary
prerequisites to that analysis, that the convection is the essential component of both the 2010 and 2014 events, in that the diabatic heating is the dominant forcing 
term in the omega equation. We also show that the magnitude of the vertical motion directly forced by orographic lifting
is consistently larger than that of the synoptic forcing due to the upper-level 
disturbances, which suggests that the orographic forcing is the more important influence on the convection.

In the rest of this study we focus on the extreme rainfall events of September 4-17, 2014, and on the two three-day extreme rainfall events of July 20-23 and 27-30, 2010, referred to hereafter as the first and second event of 2010, respectively. The data that are used are discussed in section \ref{Data}, a description of the three flood events and the moisture transport is given in section \ref{FloodDesc}, diagnoses of the influences on vertical motion using the quasi-geostrophic omega equation are given in section \ref{qgOmega}, analysis of a longer historical record to place the 2010 and 2014 events in context is in section 5, and concluding remarks are given in section \ref{Conclusion}.

\section{Data} \label{Data}
The ERA-Interim reanalysis dataset \citep{Dee11} was used for the analysis discussed in this paper.  Most of the ERA-Interim fields that were used have a 6-hourly temporal resolution and a spatial resolution of 0.7$^\circ$.  The ERA-Interim precipitation field used is the short-range ECMWF forecast, which is available at 12-hourly temporal resolution.  The ERA-Interim precipitation was  also compared to the 3B42 Tropical Rainfall Measuring Mission (TRMM) precipitation data, which has a 3-hourly temporal resolution and a 0.25$^\circ$ spatial resolution \citep{Huffman01,Huffman07}.

A flood domain used for area-averaged time-series calculations was defined as the region from 70$^\circ$ to 77$^\circ$ longitude and 30$^\circ$ to 37$^\circ$ latitude.  This region can be seen in figure \ref{topoMap}.

\section{Description of flood events} \label{FloodDesc}
\begin{figure}
%\centering
\includegraphics[width=30pc,angle=0]{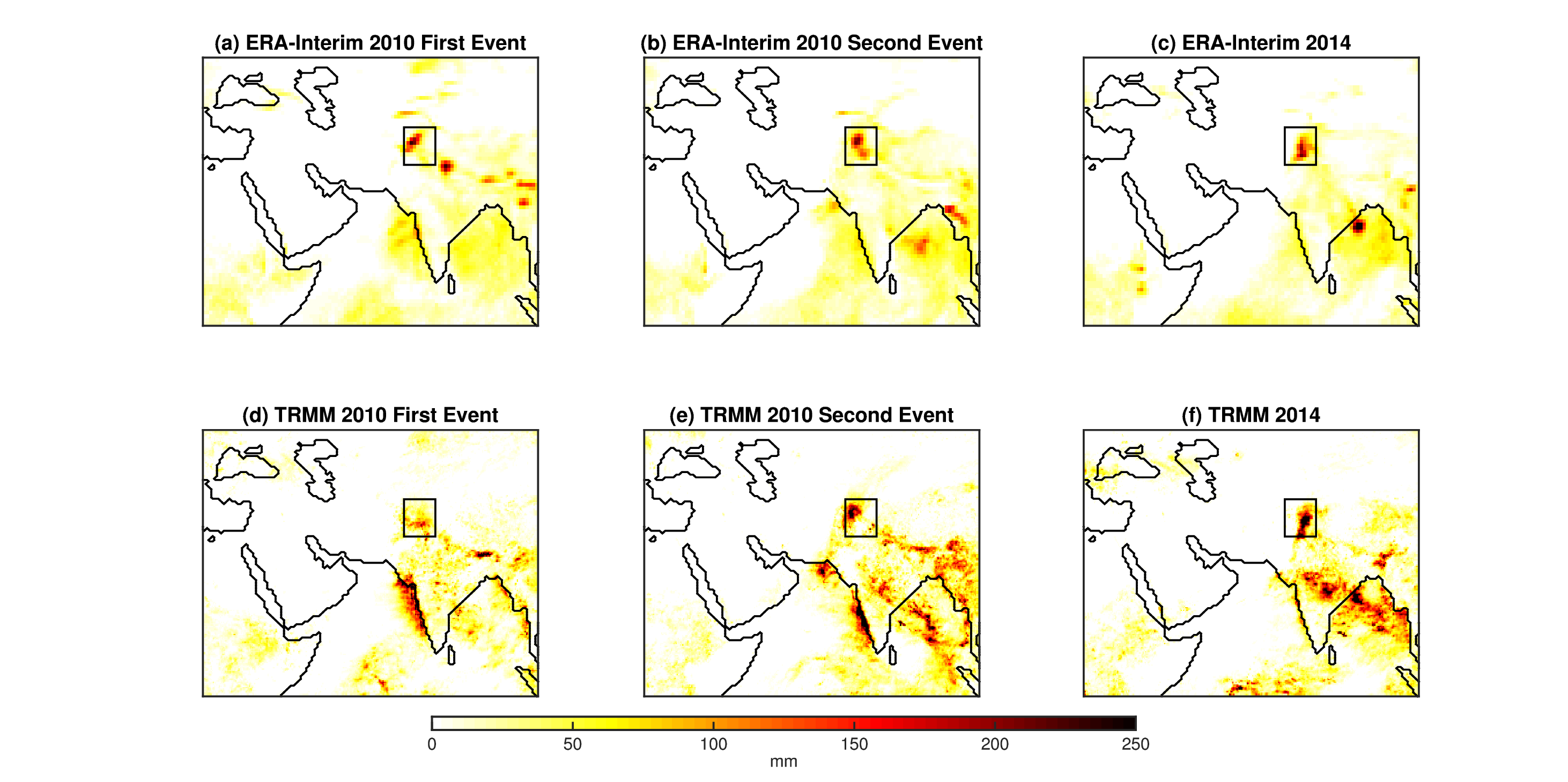}\\ 
\caption{Three-day accumulated precipitation.  (a)-(c) are for ERA-Interim and (d)-(f) is for TRMM.  (a) and (d) are for the 2010 First event: 07-20-10 to 07-23-10.  (b) and (e) are for the 2010 Second Event: 07-27-10 to 07-30-10.  (c) and (f) are for the 2014 Event: 09-04-14 to 09-07-14.  The black rectangles define the domain used for time series calculations.}
   \label{mapPrecip}
\end{figure}

Figure \ref{mapPrecip} shows the three-day accumulated precipitation of both 2010 events and the 2014 event for both the ERA-Interim and TRMM datasets.  Both the ERA-Interim and TRMM datasets show extreme rainfall totals in the respective flood domains for all three events and there is relatively good spatial agreement between the two datasets.  One discrepancy is that TRMM shows less rainfall than does ERA-Interim for the 2010 first event.  The maximum total rainfall for a single grid point in the flood domain for the 2010 first event was 196 mm and 219 mm from ERA-Interim and TRMM, respectively; the maximum total rainfall for a single grid point for the 2010 second event was 212 mm and 310 mm; and the maximum total rainfall for a single grid point in the flood domain for the 2014 event was 212 mm and 430 mm.  The differences in the maxima between the two datasets may be due in part to the TRMM data's higher spatial resolution than the ERA-Interim data.

\subsection{2010 first event: 07-20-10 to 07-23-10}
Figure \ref{timeSeriesCombine}(a)  
shows area-averaged time series of precipitation and column precipitable water during July 2010.  Prior to the first
precipitation event, a heat low was present
to the northwest of Pakistan \citep{Martius13}.  This region of low pressure can be seen in the 500-hPa height anomalies shown in figure \ref{500mbHeightAnom_withPW}(a) along with a high-pressure anomaly located over the Bay of Bengal.  These low and high pressures led to southwesterlies that transported moisture from the Arabian Sea into the flood region.  The precipitation event commenced on July 20th  and the precipitable water rose above 35 mm.

\subsection{2010 second event: 07-27-10 to 07-30-10}
Between July 25 and 28, a low pressure system traveled from the Bay of Bengal westward across India to the Arabian Sea, while a weaker low pressure system formed over the Bay of Bengal and an extreme high pressure anomaly was located over the Tibetan Plateau and northern India \citep{Houze11}.
These systems (seen in figures \ref{500mbHeightAnom_withPW}(b) and \ref{500mbHeight2010}) provided a very strong pressure gradient which resulted in southeasterly flow that transported moisture from the Bay of Bengal into the flood region \citep{Martius13,Galarneau12}).
These large pressure gradients (as well as the pressure gradients in the first 2010 event) are very unusual in this region.  
\cite{Romatschke11} found that for typical rainstorms in the western Himalayan foothills, dry air advected into the region from the Afghan Plateau
surrounds intense convective clouds and does not permit their growth into larger storms with large precipitation areas. This stands in contrast to this event, 
where the large pressure gradient brought an extreme amount of moisture into the region and allowed for very large storms to develop.
As seen in figure \ref{timeSeriesCombine}(a),  the strong pressure gradient resulted in a dramatic increase in area-average precipitable water to 45 mm.  
The very large amount of precipitable water can also be seen in figure \ref{500mbHeightAnom_withPW}(e).

\subsection{2014 event: 09-04-14 to 09-07-14}
\begin{figure}
%\centering
\includegraphics[width=30pc,angle=0]{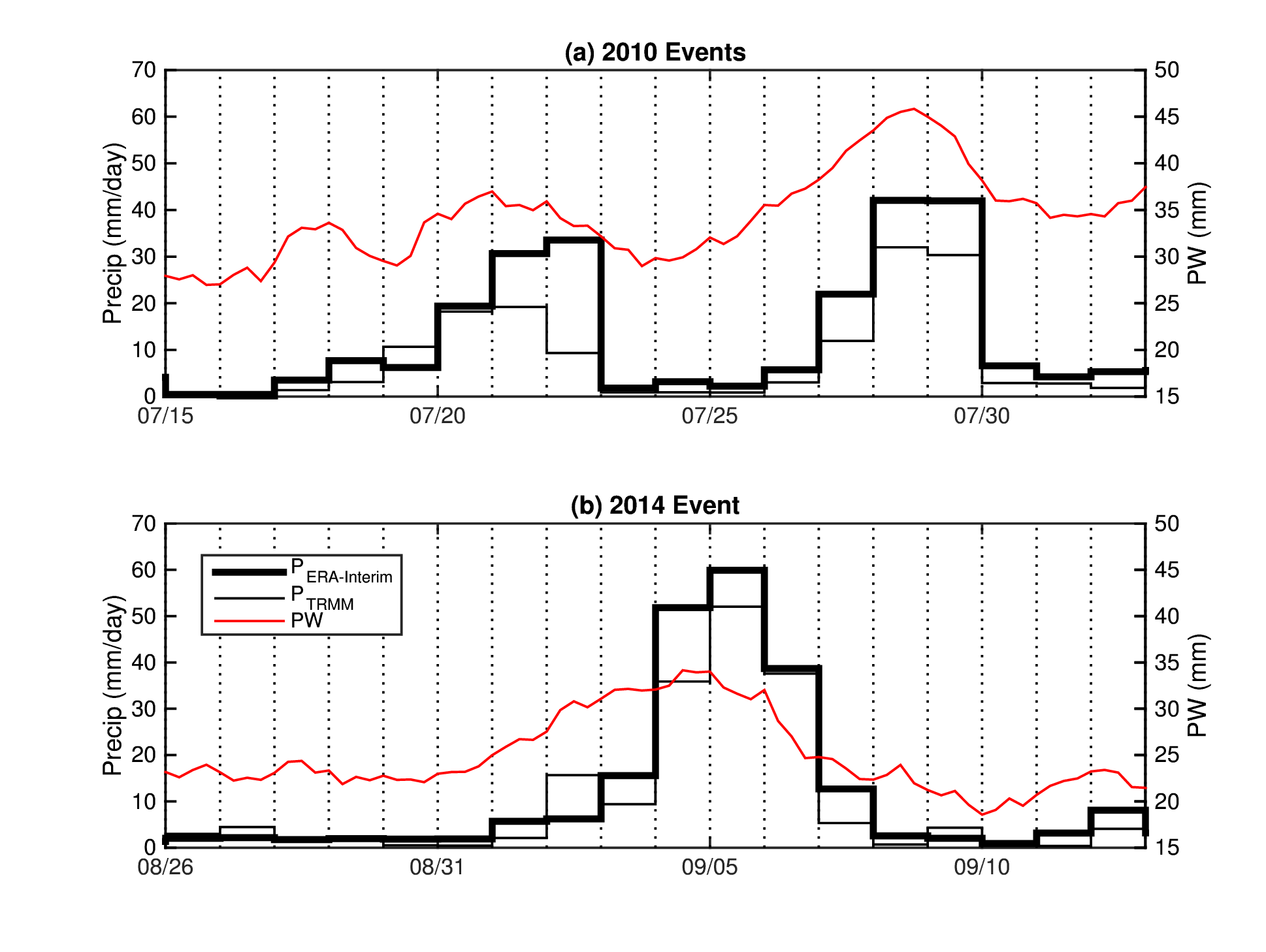}\\ 
\caption{Time series of the area averaged ERA-Interim precipitation, TRMM precipitation, and precipitable water. (a) 2010 events, (b) 2014 event.}
  \label{timeSeriesCombine}
\end{figure}

Figure \ref{timeSeriesCombine}(b) shows area-averaged time series of precipitation and precipitable water during September 2014.  Similar to the 2010 events, the evolution of precipitable water shows pre-moistening for several days before the onset of rainfall, with the precipitable water peaking during the middle of the precipitation events and then decreasing.

\begin{figure}
%\centering
\includegraphics[width=30pc,angle=0]{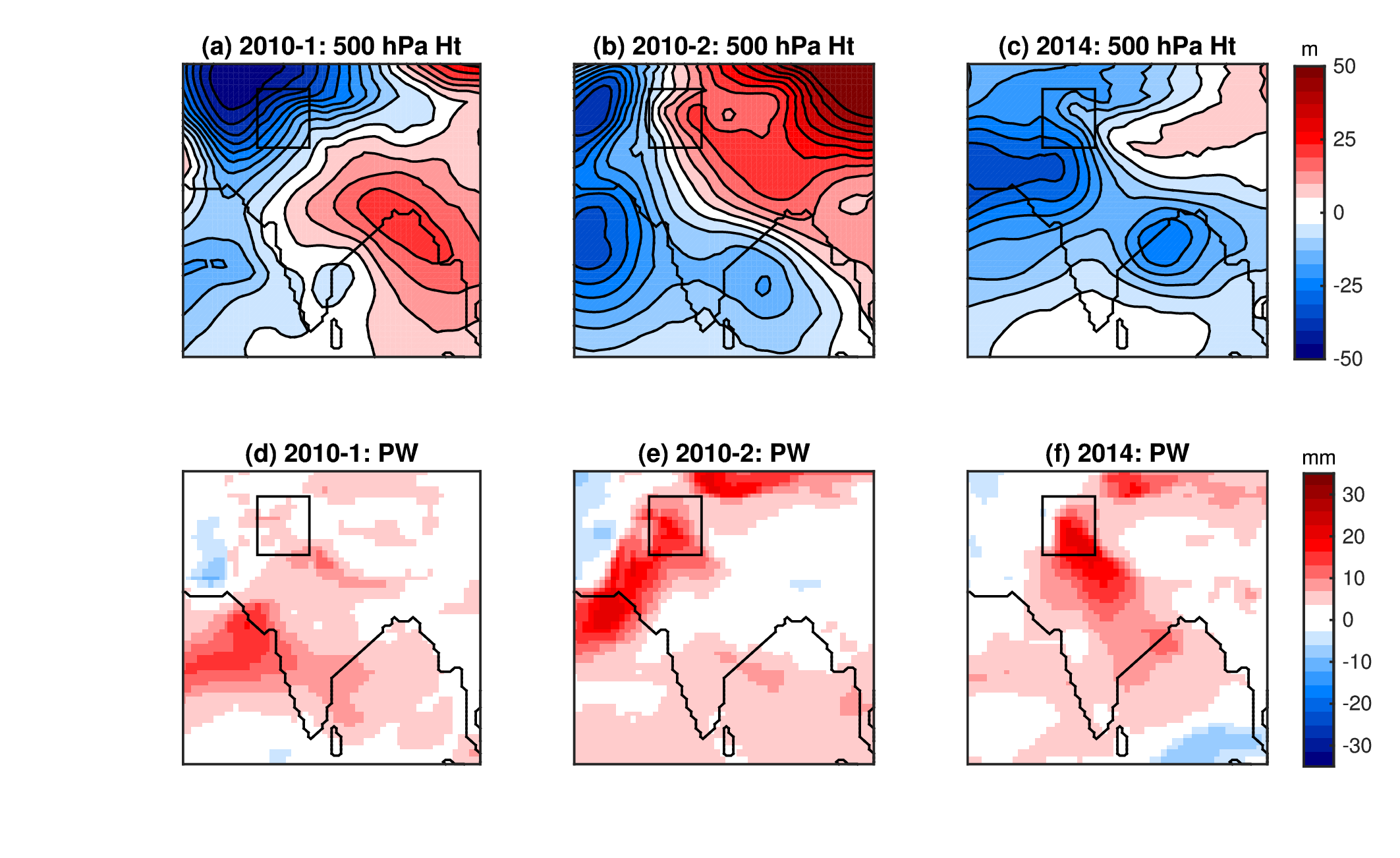}\\ 
\caption{Three day average height anomalies of the 500-hPa surface (top) and precipitable water anomalies (bottom).  (a) and (d) 2010 First Event: 07-20-10 to 07-23-10.  (b) and (e) 2010 Second Event: 07-27-10 to 07-30-10.  (c) and (f) 2014 Event: 09-04-14 to 09-07-14.}
  \label{500mbHeightAnom_withPW}
\end{figure}

Figure \ref{500mbHeightAnom_withPW}(f) shows that there was anomalously high precipitable water over all of Pakistan, which is similar, though less intense, to the state
during the second event of 2010 (figure \ref{500mbHeightAnom_withPW}(e)).

\begin{figure}
%\centering
\includegraphics[width=30pc,angle=0]{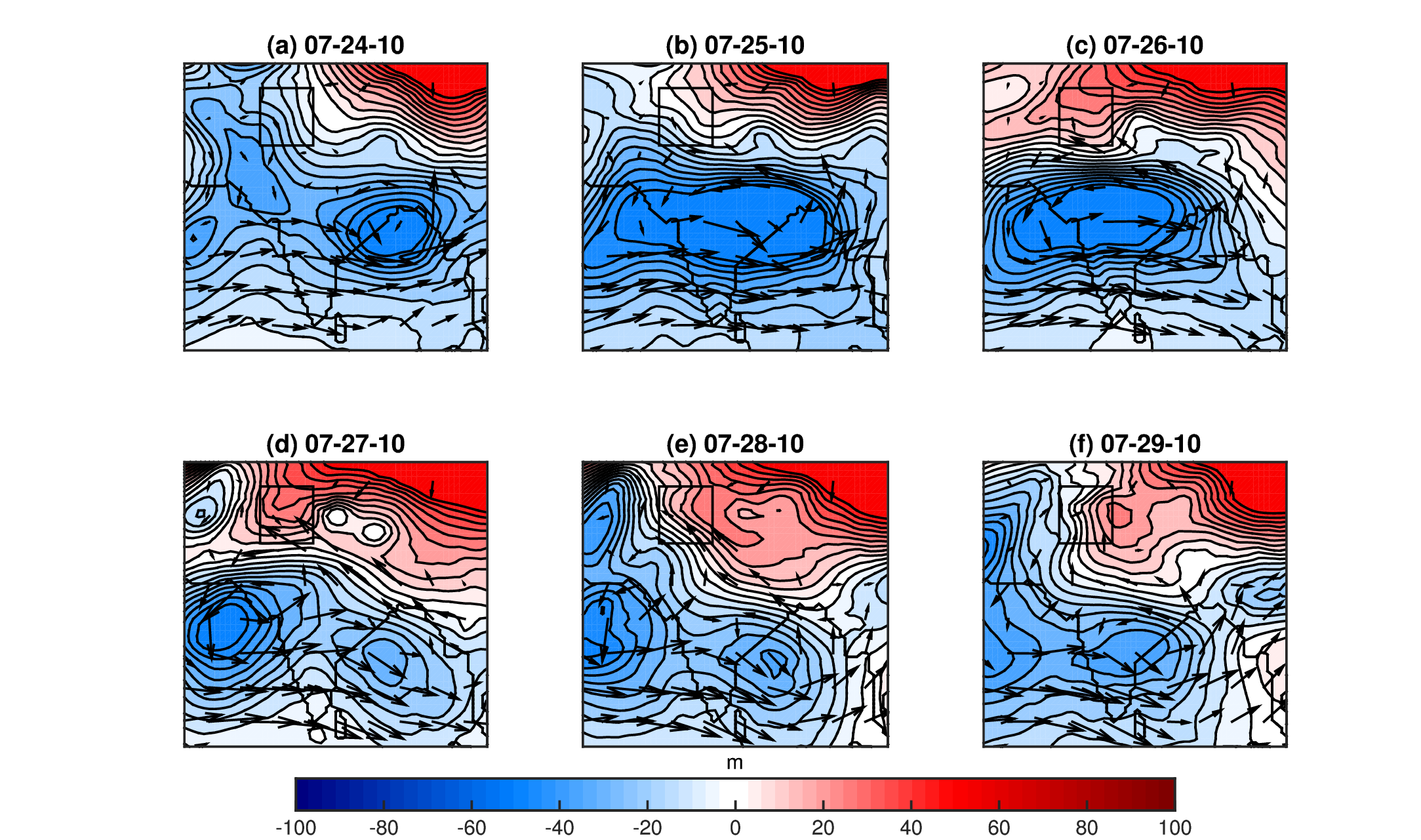}\\ 
\caption{Sequence of maps showing the evolution of one day average height anomalies of the 500-hPa surface 
and one day average moisture flux at 700-hPa (arrows)  from 07-24-10 to 07-29-10.}
  \label{500mbHeight2010}
\end{figure}

\begin{figure}
%\centering
\includegraphics[width=30pc,angle=0]{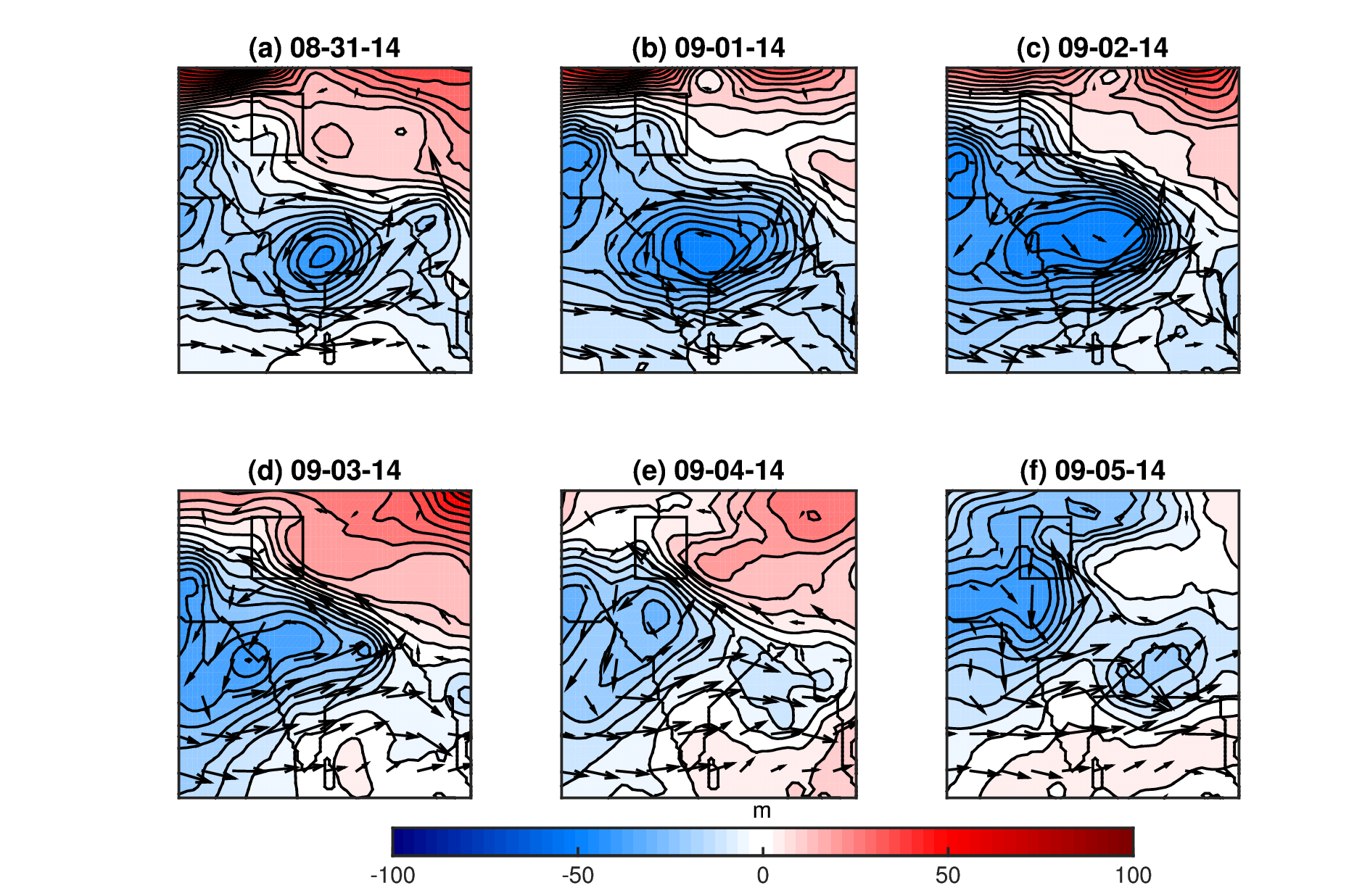}\\ 
\caption{Sequence of maps showing the evolution of one day average height anomalies of the 500-hPa surface (colors) and one day average moisture flux at 700-hPa (arrows)  from 08-31-14 to 09-05-14.}
  \label{500mbHeight2014}
\end{figure}

Figure \ref{500mbHeightAnom_withPW} (a-c) also shows large variability in the height field.  The first event of 2010 has a northwest-southeast dipole, the second event of 2010 has a southwest-northeast dipole, and the 2014 event has a west-east dipole.  These dipoles are consistent with the low level flow which transported moisture into the flood region: the moisture for the first event of 2010 came form the Arabian sea, while the moisture for the second event of 2010 and the 2014 events came from the Bay of Bengal (figures \ref{500mbHeight2010} and \ref{500mbHeight2014}).

Figure \ref{500mbHeight2014} shows the time evolution of the 500 hPa height field anomaly along with the 700 hPa moisture flux.  From 08-31-14 to 09-03-14 there was a similar pressure configuration to the second event of 2010, with a strong depression moving westward over India and a high pressure anomaly
over the Tibetan Plateau, which resulted in southeasterlies transporting moisture from the Bay of Bengal into India and Pakistan and the
area-average precipitable water in the flood region increasing from roughly 20 mm to 35 mm (figure \ref{timeSeriesCombine}(b)).  The similarity of the evolution of this flow pattern can be seen in comparing figures \ref{500mbHeight2014} and \ref{500mbHeight2010}.  The Tibetan anti-cyclone dissipated in the 2014 event instead of persisting like the 2010 event, but the position and timing of the transit of the low pressure across India is very similar.  Subsequent to the low pressure transit, the high pressure anomaly weakened and a new depression formed over the Bay of Bengal.

\section{Quasi-geostrophic Omega Equation} \label{qgOmega}
The quasi-geostrophic omega equation is a diagnostic equation for the quasi-geostrophic pressure vertical velocity, $\omega$:
\begin{align}
\left( \nabla^2 + \frac{f_0^2}{\sigma} \frac{\partial^2}{\partial p^2} \right) \omega =  
& \frac{f_0}{\sigma} \frac{\partial}{\partial p} \left[ v_g \cdot \nabla \left( \frac{1}{f_0} \nabla^2 \Phi + f \right) \right]  \nonumber \\
& + \frac{1}{\sigma} \nabla^2 \left[ v_g \cdot \nabla \left( - \frac{\partial \Phi}{\partial p} \right) \right] \nonumber \\
& - \frac{\kappa}{\sigma p} \nabla^2 J 
\label{omegaEquation}
\end{align}
where $f$ is the coriolis parameter, $f_0$ here is $f$ at a latitude of 33.5$^{\circ}$, $\sigma=-\frac{RT_0}{p}\frac{d ln \theta_0}{dp}$ is the static stability, $R$ is the gas constant,
$T$ is temperature, $\theta=T\left( \frac{p_0}{p} \right) ^\kappa$ is potential temperature, $v_g=\frac{1}{f_0}\hat{k} \times \nabla \Phi$ is the geostrophic velocity, $\Phi$ is the geopotential,
$\kappa=\frac{R}{c_p}$, $c_p$ is the specific heat, $\nabla ^2$ is the horizontal Laplacian,
and $J$ is the heating rate.  
The first term on the right hand side of Eq.
(\ref{omegaEquation}) represents differential vorticity advection, the second term is the horizontal
Laplacian of temperature advection, and the third term is the horizontal Laplacian of diabatic
heating.  We refer to the first two terms as contributions from the large-scale circulation, or as synoptic forcing.

The right hand side of Eq. (\ref{omegaEquation}) can be computed solely from the geopotential, $\Phi$, and the heating rate, $J$.  
$\Phi$ is available in the ERA-Interim reanalysis, but $J$ is not.  $J$, therefore,  is diagnosed from the temperature tendency equation:
\begin{equation}
\left( \frac{\partial}{\partial t} + v_g \cdot \nabla \right) \left( - \frac{\partial \Phi}{\partial p} \right) - \sigma \tilde{\omega} 
= \frac{\kappa J}{p}
\label{temperatureEquation}
\end{equation}
where $\tilde{\omega}$ is the actual pressure vertical velocity, which need not be identical to the quasi-geostrophic
one, $\omega$, derived from (\ref{omegaEquation}). Here we obtain $\tilde{\omega}$ from the ERA-Interim reanalysis. 

Equation (\ref{omegaEquation}) was inverted for the three flood events between 550 hPa and 175 hPa using Stone's method, also known as the strongly implicit procedure, \citep{Stone68}.
In order to decompose $\omega$, the inversion is first carried out with only the first two forcing terms on the right hand side of equation \ref{omegaEquation} (referred to as the contribution of the synoptic advection terms) and a lower boundary condition of $\omega$ set to zero.  Secondly, the inversion is carried out with only the third term on the right hand side of equation \ref{omegaEquation} (referred to as contribution of the convective heating term) and a lower boundary condition of $\omega$ set to zero.  Thirdly, the inversion is carried out with no forcing terms on the right hand side of equation \ref{omegaEquation} (referred to as the contribution of the boundary condition) and a lower boundary condition of $\omega$ set to a topographic omega, $\omega_{qg}(p_s)$, which is described below.

The lower boundary of 550 hPa was chosen because
the flood regions have a very large topographic gradient inside them which leads to time-mean surface pressures ranging from 990 hPa to
585 hPa.  Choosing a lower boundary at 550 hPa ensures that the entire domain is above the surface.\footnote{An alternative approach is detailed in the Appendix.}
  The upper boundary of 175 hPa
was chosen as a nominal tropopause.  The domain used to perform the
inversion was from 0$^\circ$ to 55$^\circ$ latitude and 25$^\circ$ and 100$^\circ$ longitude.  $\omega_{qg}$ was set to zero at the horizontal 
boundaries and at the upper boundary.  The horizontal domain size is large enough that the horizontal boundary conditions have no effect on the solution in the flood region. 

The resolution used in the inversion is the same as the resolution of the ERA-Interim data, which has horizontal resolution of 0.7$^\circ$ and vertical resolution of 25 $hPa$ in the lower troposphere and 50 $hPa$ in the mid-troposphere.

The topographic vertical velocity, $w(p_s)$, is computed using the geostrophic velocities at the surface:
\begin{equation}
w(p_s) = \boldsymbol{v}_{g}(p_s) \cdot \nabla h
\label{wTopo}
\end{equation}
where $\boldsymbol{v}_{g}(p_s)$ is the geostrophic velocity linearly interpolated to the local surface pressure, $p_s$, and
$h$ is the height of the topography.  The topographic vertical velocity is then converted to the topographic omega using the hydrostatic
approximation:
\begin{equation}
\omega(p_s) = - \rho g w(p_s)
\end{equation}

\begin{figure}
%\centering
\includegraphics[width=30pc,angle=0]{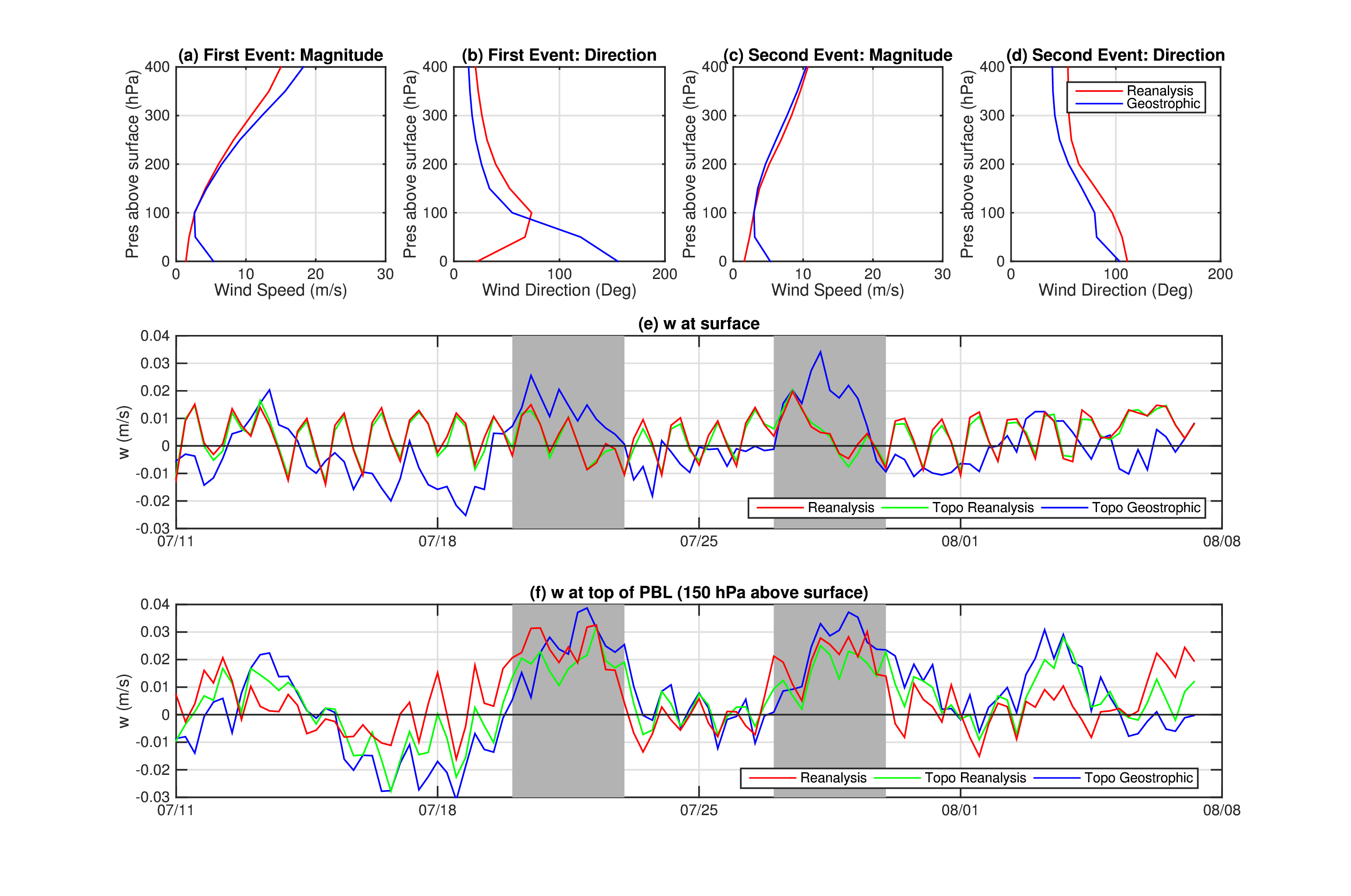}\\ 
\caption{Top row: Area-average reanalysis (red) and geostrophic (blue) velocities as a function of pressure above the surface for the peak of the first event of 2010 (07-22-10 0000), (a) and (b), and the peak of the second event of 2010 (07-29-10 0600), (c) and (d). (a) and (c) show the magnitude of the velocity and (b) and (d) show the velocity direction.  
Middle and bottom rows: reanalysis vertical velocity (blue) as well as topographic forced vertical velocity using the reanalysis wind (red) and geostrophic wind (yellow).  (e) is for the vertical velocity at the surface and (f) is for the vertical velocity at the top of the boundary layer, defined as 150 hPa above the surface. The gray bars indicate the precipitation
events. 
}
  \label{topoPBL}
\end{figure}

In the presence of a sloping lower boundary, the natural boundary condition to use is the topographically forced vertical velocity by the geostrophic wind at a level close to the surface. Here we consider the appropriate level to use for this purpose.
%ahs I don't think we need the sentence below; in QG I think actually there is a nominal notion of a PBL that should be used.
%The implicit assumption is that the geostrophic and actual velocities will be approximately equal and the subsequent topographic forced vertical velocity will match the actual vertical velocity at the surface.  
Figures \ref{topoPBL} (a)-(d) show the magnitude and direction of the reanalysis and geostrophic velocities as a function of the pressure above the surface.  While the reanalysis velocities are well approximated by the geostrophic velocities in the free troposphere, there is clearly a planetary boundary layer (PBL) near the surface where frictional effects cannot be neglected and the velocity is not geostrophic.  

Figure \ref{topoPBL} (e) shows a comparison of the reanalysis vertical velocity at the surface, the topographic forced vertical velocity by the surface reanalysis winds, and the topographic forced vertical velocity by the surface geostrophic winds.  The reanalysis vertical velocity and the topographic forced vertical velocity by the surface reanalysis winds closely match and show a strong diurnal cycle with no visible signal during the flood events.  Due to the mismatch of the reanalysis and the geostrophic winds at the surface, the topographic forced vertical velocity by the surface geostrophic winds do not match the others and is the only one to have an upward vertical velocity signal during the flood events.

Figure \ref{topoPBL} (f) shows a comparison of the reanalysis vertical velocity at the top of the PBL, the topographic forced vertical velocity by the reanalysis winds at the top of a nominal PBL, defined as a level 150 hPa above the surface, and the topographic forced vertical velocity by the geostrophic winds at the top of the PBL.  Now, all of the velocities match well and show a positive vertical velocity signal during the flood events.  Other PBL heights were tested (not shown) and a height of 150 hPa was found to deliver the best match.  It is high enough to ensure that the reanalysis winds are well approximated by the geostrophic winds and low enough to ensure that the reanalysis vertical velocity matches the topographic forced vertical velocity.  Therefore, the topographic forced vertical velocity by the geostrophic winds at a PBL height of 150 hPa was used as the lower boundary condition for the inversion of the omega equation.

$\omega(p_s)$ is defined at the top of the PBL.  This level ranges from a pressure of roughly 840 hPa to 440 hPa, so using it as the lower boundary condition at 550 hPa is somewhat artificial.  Nevertheless, it is a simple method for attempting to include the effects of orographic forcing. 

\begin{figure}
%\centering
\includegraphics[width=30pc,angle=0]{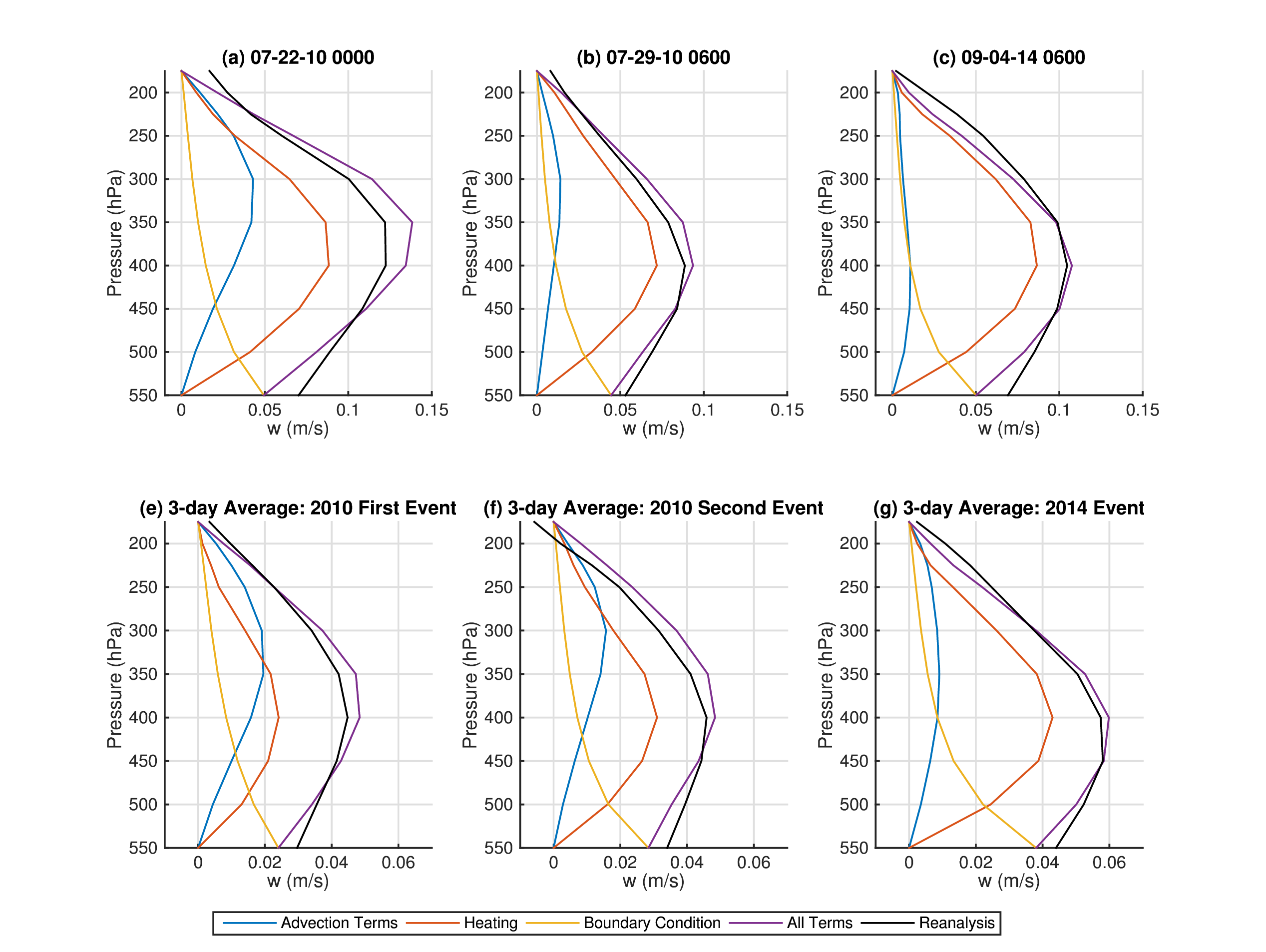}\\ 
\caption{Area average vertical velocity, w, profiles from inverting the omega equation.  Shown are the contribution from the advection terms (blue), the heating term (red), the boundary condition (yellow), the sum of all terms (purple), and the Reanalysis vertical velocity (black).   The top plots are for points-in-time and the bottom plots are three-day averages during each event.  2010 First Event: (a) 07-22-10 0000 and (e) three-day average from 07-20-10 and 07-23-14.  2010 Second Event: (b) 07-29-10 0600 and (f) three-day average from 07-20-10 and 07-23-14.
2014 Event: (c) 09-04-14 0600 and (g) three-day average from 09-04-14 to 09-07-14.}
  \label{wProfileAllEvents}
\end{figure}

Figures \ref{wProfileAllEvents}(a) and \ref{wProfileAllEvents}(e) show area-averaged profiles of $w$ from inverting the omega equation, split into the contributions from the synoptic advection terms, the convective heating term, and the boundary condition, as well as the reanalysis $w$ for the first event of 2010.  Shown are point-in-time profiles at 07-22-10 0000 (which was the time of maximum vertical velocity for this event) and a three-day average over the whole event from  07-20-10 to 07-23-14.  Both the point-in-time and time-averaged solutions match the reanalysis profile well, with the point-in-time solution overestimating slightly in the middle of the domain.  In the point-in-time solution, while the contribution of the heating term is greatest, the advection terms contribute significantly, with the advection terms contribution being roughly half of the heating term contribution at mid-levels.  For the time-averaged solution, the advection terms' net contribution is of the same order as the heating term contribution.  The effect of the topographic lower boundary condition decays with height, as expected, and the boundary condition underestimates the reanalysis vertical velocity at 550 hPa.

Similarly, figures \ref{wProfileAllEvents}(b) and \ref{wProfileAllEvents}(f)  show area-averaged profiles of $w$ from inverting the omega equation as well as the reanalysis $w$ for the second event of 2010.  Again, shown are point-in-time profiles at 07-29-10 0600 (which was the time of maximum vertical velocity for this event) and a three-day average over the whole event from  07-27-10 to 07-30-14.  Both the point-in-time and three-day average solutions again match the reanalysis profiles closely.  In contrast to the first event, the heating term dominates to a larger extent.  In particular, the contribution of the advection terms in the point-in-time solution is much weaker than in the first event.

Figures \ref{wProfileAllEvents} (c) and \ref{wProfileAllEvents} (g) show area-averaged profiles of $w$ from inverting the omega equation as well as the reanalysis $w$ for the 2014 event.  Shown are point-in-time profiles at 09-05-14 0600 (which was the time of maximum vertical velocity for this event) and a three-day average over the whole event from  09-04-14 to 09-07-14.  As with the 2010 events, the overall agreement between the total solution and the reanalysis profile is quite good.  
The heating term completely dominates in both the point-in-time and time-averaged solutions, which is similar to the second event of 2010 but to an even greater extent.

For all three events, the magnitude of the lower boundary condition on the vertical velocity is of the same order as the interior maxima associated with the heating term.  This closeness in magnitude suggests that the topographic forced lifting of the lower boundary condition may be a significant trigger for the convective heating. Further investigation of this inference using cloud-resolving model simulations is presented by Nie et al. (2016).

\begin{figure}
%\centering
t\includegraphics[width=30pc,angle=0]{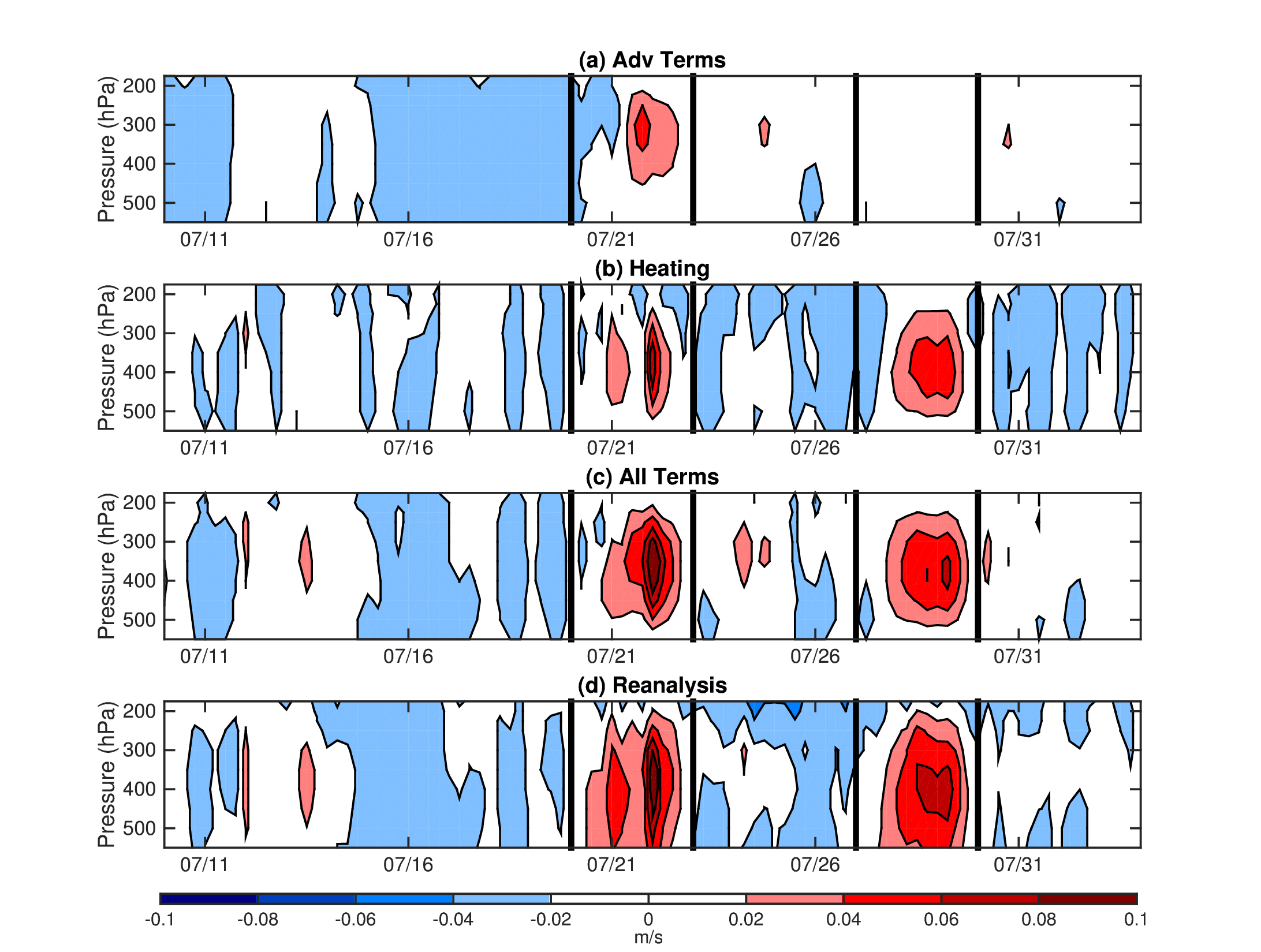}\\ 
\caption{2010 event.  Area average vertical velocity through time from inverting the omega equation.  Shown are the contribution from (a) the advection terms , (b) the heating term, (c) the sum of all terms (without the boundary condition), and (d) the Reanalysis vertical velocity.  The black lines bound the time of the flood events.}
  \label{wProfileTime2010}
\end{figure}

\begin{figure}
%\centering
\includegraphics[width=30pc,angle=0]{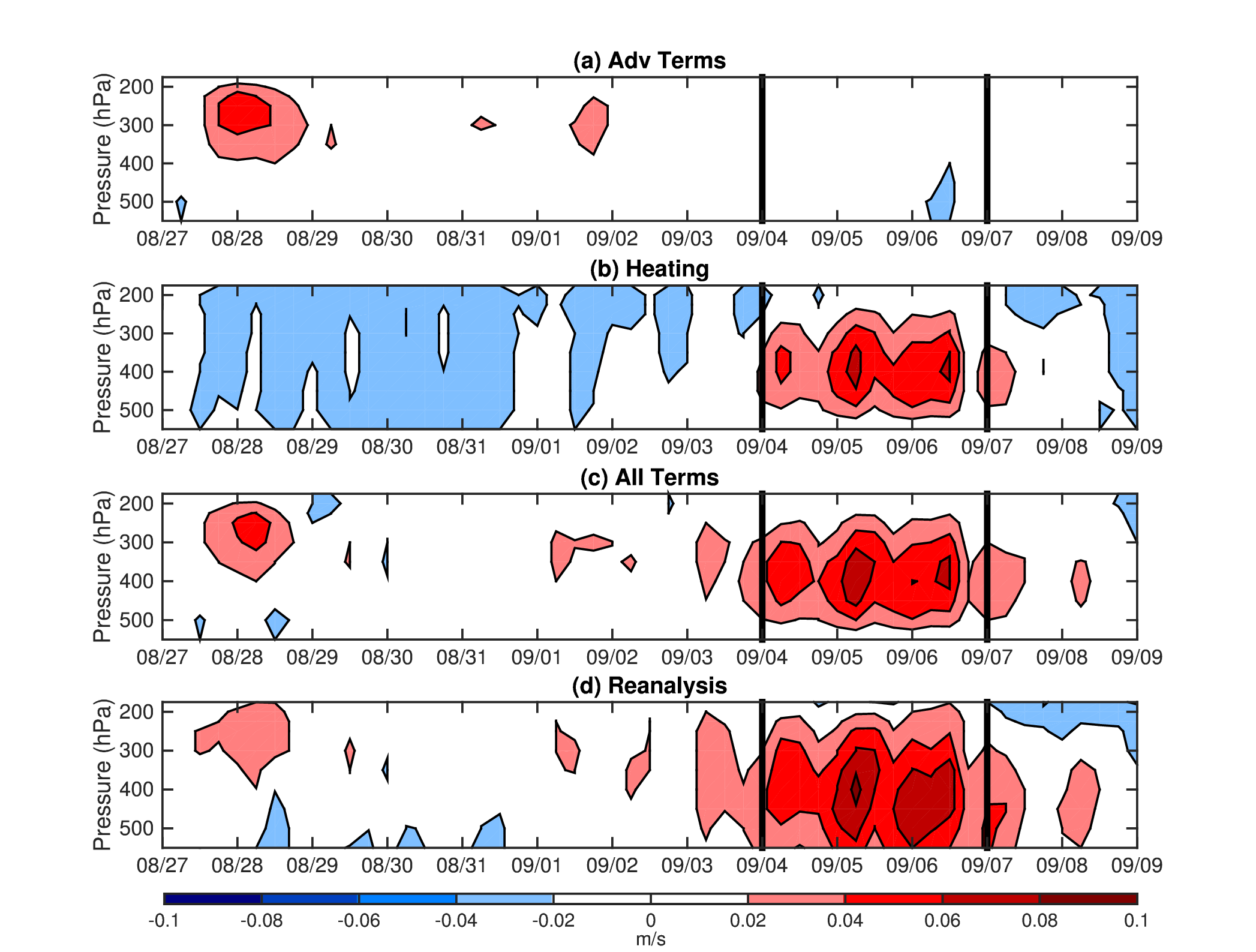}\\ 
\caption{2014 event.  Area average vertical velocity through time from inverting the omega equation.  Shown are the contribution from (a) the advection terms , (b) the heating term, (c) the sum of all terms (without the boundary condition), and (d) the Reanalysis vertical velocity.  The black lines bound the time of the flood event.}
  \label{wProfileTime2014}
\end{figure}

Figure \ref{wProfileTime2010} shows the time evolution spanning both 2010 events of the $w$ profiles from inverting the omega equation and the reanalysis.  Shown are the sum of the contributions from the two advection terms (which captures the effect of the large scale circulation), the contribution of the heating term, the sum of all terms (excluding the boundary condition), and the reanalysis.  The total solution matches the reanalysis $w$ well for both events, with the exception of the first weak peak of the first event which the solution underestimates.  As seen in the previous two figures, the advection terms, and thus the large scale circulation,  have a much larger effect during the first event than the second event, which was also found by \cite{Martius13}.  
Figure \ref{wProfileTime2014} shows the time evolution of the $w$ profiles from inverting the omega equation and the reanalysis for the 2014 event.  Like the second event of 2010, there is no discernible signal in the advection terms during the flood event, with all of the vertical velocity being forced by heating.

\begin{figure}
%\centering
\includegraphics[width=20pc,angle=0]{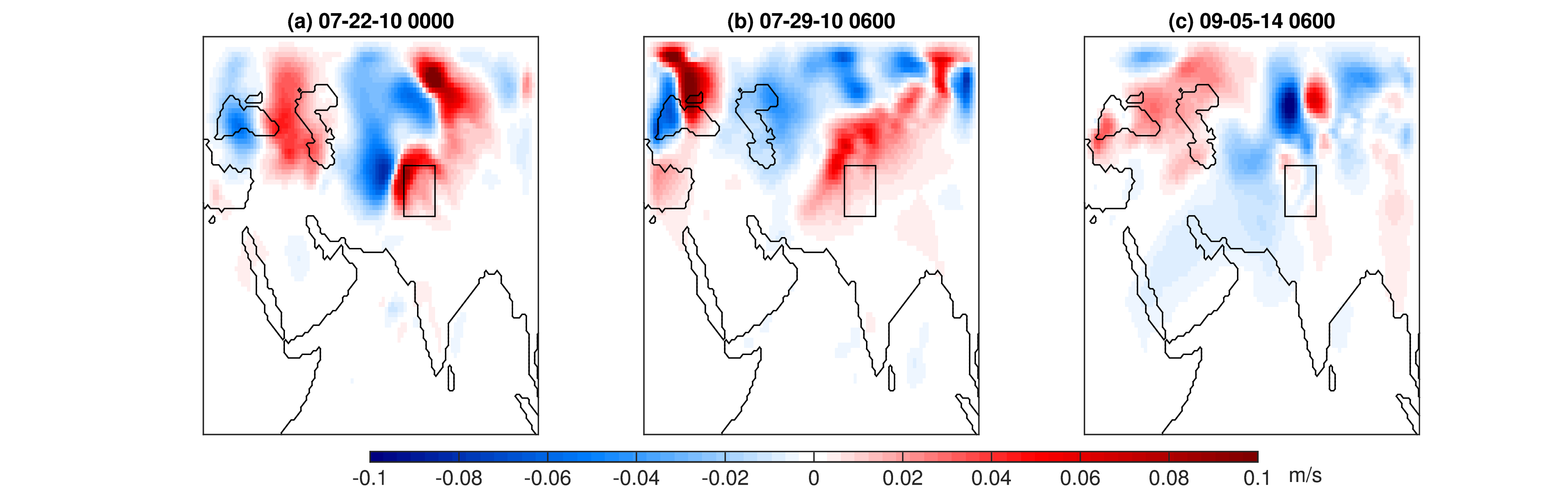}\\ 
\caption{Map of 350 hPa vertical velocity from inverting the omega equation.  Only contributions from the advection terms are included.  (a) 07-22-10 0000, (b) 07-29-10 0600, and (c) 09-05-14 0600. }
  \label{wMaps}
\end{figure}

Figure \ref{wMaps} shows maps of the large-scale circulation forced $w$ at 350 hPa during the peaks of the three events.  As expected, the solution is more active poleward of the flood regions where there are larger wind velocities due to the jet stream.  As seen before, only the first event of 2010 has significant forced lifting by the circulation.

\section{Historical comparison}
The above analysis suggests that the amount of precipitable water and the topographic forced vertical velocity were the dominant causes of the three flood events.  It is useful to determine if the relationship between these forcings and extreme amounts of precipitation in this region found for the 2010 and 2014 events also exists historically.  This will also reveal how often the levels of precipitable water, topographic forced lifting, and precipitation rate seen during the flood events occur in this region.   To this end, the monsoon seasons were examined from 2004 through 2014.

\begin{figure}
%\centering
\includegraphics[width=30pc,angle=0]{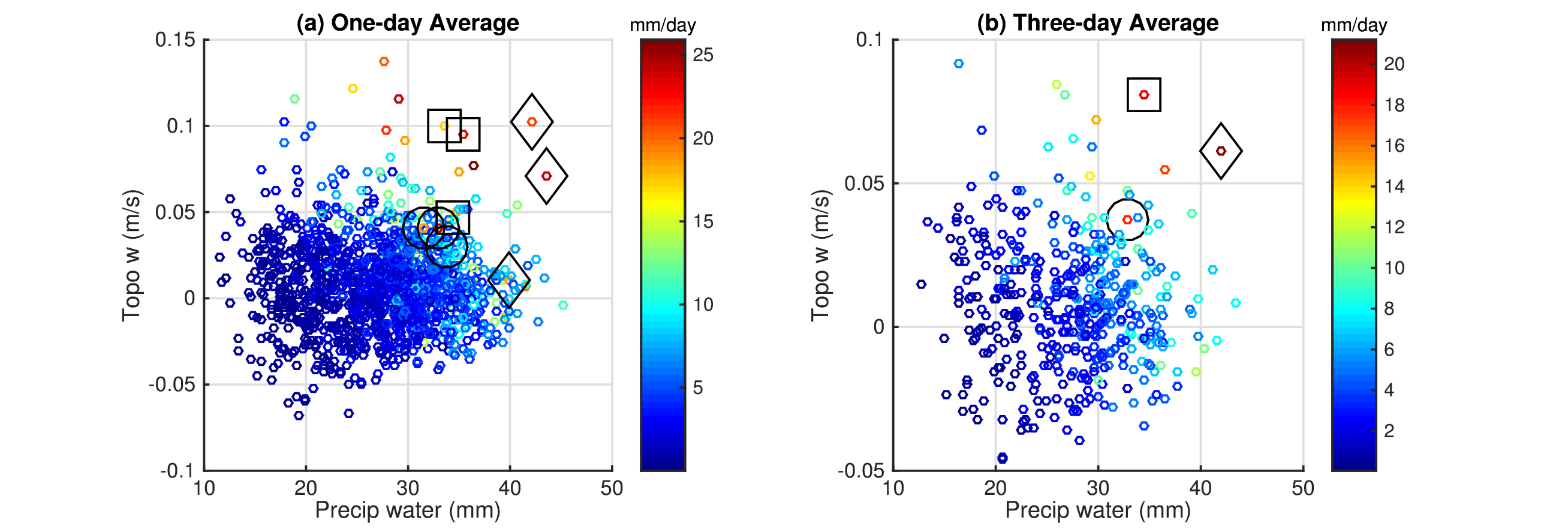}\\ 
\caption{Average precipitable water and topographic forced vertical velocity.  The colors are precipitation rate in mm/day.  The first event of 2010 is shown with black squares, the second event of 2010 is shown with black diamonds, and the 2014 event is shown with black circles.  Time range is June - September 2004-2014.    (a) One-day averages, (b) Three-day averages.  }
  \label{scatterBoth}
\end{figure}

Figure \ref{scatterBoth}(a) shows flood-domain averaged precipitation rate for one-day averaged time periods for June though September of the years 2004 though 2014 as a function of the amount of precipitable water and the topographic forced vertical velocity.  The three flood events examined in this study are indicated with black outlines.  There is a clear relationship between the precipitation and the other two variables, with days with high levels of precipitable water and topographic forcing corresponding to days with high levels of precipitation.

Figure \ref{scatterBoth}(b) is similar to figure \ref{scatterBoth}(b) but with three-day averages instead of one-day averages.  There is again a clear relationship between the precipitation and the other two variables, with the two events of 2010 being particularly extreme in all variables and the 2014 event somewhat weaker.
%ahs I don't think we need to say this, and could imagine it getting us in trouble Previous studies \citep{Webster11,Galarneau12,Rasmussen15} using the ECMWF Ensemble Prediction System (EPS) have shown the ability to accurately forecast that conditions that favor floods of this type in northeast Pakistan with up to 1-2 weeks lead time.  The results shown here that extreme precipitation in this region is chiefly associated with moisture influx and topographic lifting could be a physical reason for the ability of the forecasts to predict these types of events.

\section{Conclusion} \label{Conclusion}
Devastating floods in northeast Pakistan and north India occurred in July 2010 and September 2014.  These floods were
the result of three extreme precipitation events (two in July 2010 and one in September 2014).  This study has compared the events and attempted to describe the factors that supported the events, specifically the circulation that brought moisture into the region and the driving factors of the upward motion that led to the precipitation.

The vertical motion was analyzed through inverting the quasi-geostrophic omega equation, which allowed for the contribution of the large-scale circulation forcing to be quantified.   The omega equation solutions matched the reanalysis vertical motion fairly well for all three events.  The topographic forced vertical motion was also calculated using the geostrophic winds at the top of the PBL.

All three events had anomalously high amounts of moisture in the region due to the configuration of the flow field that transported moisture into the region from either the Bay of Bengal or the Arabian Sea.  Large amounts of moisture can induce unusually extreme precipitation events from the orographic forcing due to the large topographic gradient of the Himalayas \citep{Houze11}.

It is found that while all three events had large pre-moistening prior to the onset of rainfall, the 500 hPa height surfaces and moisture sources show differences.  The specific flow pattern and moisture source in the 2014 event were similar to those in the second event of 2010.  During these events, there was a strong high pressure over the Tibetan plateau that, along with a depression that traveled from the Bay of Bengal northwest across India, caused southeasterlies which transported moisture from the Bay of Bengal onto the continent and into the flood region.  In contrast, during the first 2010 event, a high pressure in the Bay of Bengal and a low pressure northeast of Pakistan led to southwesterlies and moisture transport northeastward from the Arabian Sea.  

In contrast to the synoptic analysis, all three events show quite similar features when focusing only on the local air column and performing the quasi-geostrophic omega decomposition.  The heating term is dominant, which explains the need for pre-moistening.  The topographic forced upward motion is found to be the main triggering factor of the convection, with the synoptic forcing due to vorticity and temperature advection being relatively weak.  
This is consistent with earlier work of \cite{Sanders84} which computed the quasi-geostrophic $\omega$ for a monsoon depression in 1979 and found that the direct synoptic forcing of $\omega$ was only marginally detectible.

The conclusion drawn from these three events are also supported by studying historical data of this region.   An examination of the monsoon season over eleven years demonstrates that topographic forced ascent and moisture content are the main drivers of extreme precipitation.  By identifying common features among those extreme events, the conclusion of this study improves our understanding of the dynamics of these events, and highlights the two most relevant environmental factors which regulate their occurrence.

%\ack{The authors thank Maxi B\"{o}ttcher for providing an implementation of the Stone method to invert the omega equation. This work was supported by NASA grants NNX12AB87A and NNX15AJ05A, an AXA Award from the AXA Research Fund to AHS, and a Lamont Postdoctoral Fellowship from the Lamont-Doherty Earth Observatory to JN. 

%%% End of body of article:

%%%%%%%%%%%%%%%%%%%%%%%%%%%%%%%%
%% Optional Appendix goes here
%
% \appendix resets counters and redefines section heads
% but doesn't print anything.
% After typing \appendix
%
%\section{Here Is Appendix Title}
% will show
% Appendix A: Here Is Appendix Title
%
\section*{Appendix}
The omega equation inversion described in section \ref{qgOmega} involved a complication due to the fact that the flood region contains very tall mountains that extend up to 585 hPa.   In order to ensure that the entire domain is above the surface, the lower boundary was set to 550 hPa and the topographic vertical velocity was applied at this level.

An alternative approach that is considered here is to use a lower boundary that is below the surface in some locations.
This involves computing a solution below the surface at some locations and setting the forcing terms in Eq. (\ref{omegaEquation}) to zero below the surface.  It was shown in section \ref{qgOmega} that the geostrophic velocities at 150 hPa above the surface work well when computing the topographic vertical velocity.   The alternative approach sets the lower boundary at 700 hPa because this is the level that is roughly 150 hPa above the mean surface pressure in the flood region.  With this lower boundary, the boundary condition was defined at each location as either the reanalysis vertical velocity at 700 hPa (if the surface is below 700 hPa) or at the surface (if the surface is above 700 hPa). 

\begin{figure}
%\centering
\includegraphics[width=30pc,angle=0]{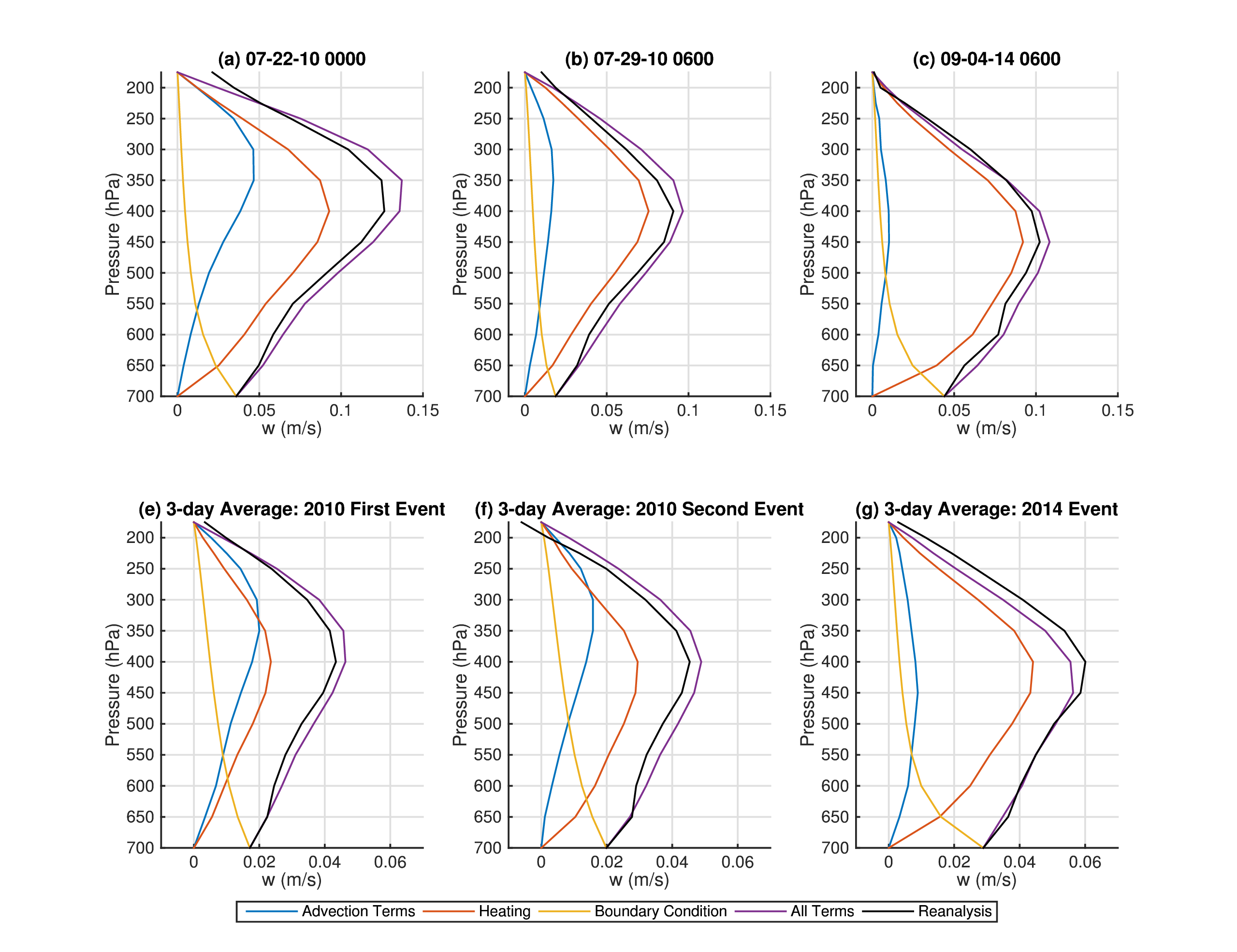}\\ 
\caption{Same as figure \ref{wProfileAllEvents} except using a lower boundary at 700 hPa as discussed in the Appendix.}
  \label{wProfileAllEvents_700}
\end{figure}

Figure \ref{wProfileAllEvents_700} is the same as figure \ref{wProfileAllEvents} but using this alternative approach with the lower boundary at 700 hPa.  
The profiles show area-averages in the flood region, where only locations above the surface are used in the averaging at each level.
Both approaches show that the solutions match the reanalysis velocity profiles well.  Both approaches also show that the heating term is of much more important than the advection terms for all events except the first event of 2010.

%%%%%%%%%%%%%%%%%%%%%%%%%%%%%%%%%%%%%%%%%%%%%%%%%%%%%%%%%%%%%%%%
%
% Optional Glossary or Notation section, goes here
%
%%%%%%%%%%%%%%
% Glossary is only allowed in Reviews of Geophysics
% \section*{Glossary}
% \paragraph{Term}
% Term Definition here
%
%%%%%%%%%%%%%%
% Notation -- End each entry with a period.
% \begin{notation}
% Term & definition.\\
% Second term & second definition.\\
% \end{notation}
%%%%%%%%%%%%%%%%%%%%%%%%%%%%%%%%%%%%%%%%%%%%%%%%%%%%%%%%%%%%%%%%
%
%  ACKNOWLEDGMENTS

\section*{Acknowledgments}
The authors thank Maxi B\"{o}ttcher for providing an implementation of the Stone method to invert the omega equation. This work was supported by NASA grants NNX12AB87A and NNX15AJ05A, an AXA Award from the AXA Research Fund to AHS, and a Lamont Postdoctoral Fellowship from the Lamont-Doherty Earth Observatory to JN. 

%% ------------------------------------------------------------------------ %%
%%  REFERENCE LIST AND TEXT CITATIONS
%
% Either type in your references using
% \begin{thebibliography}{}
% \bibitem{}
% Text
% \end{thebibliography}
%
% Or,
%
% If you use BiBTeX for your references, please use the agufull08.bst file (available at % ftp://ftp.agu.org/journals/latex/journals/Manuscript-Preparation/) to produce your .bbl
% file and copy the contents into your paper here.
%
% Follow these steps:
% 1. Run LaTeX on your LaTeX file.
%
% 2. Make sure the bibliography style appears as \bibliographystyle{agufull08}. Run BiBTeX on your LaTeX
% file.
%
% 3. Open the new .bbl file containing the reference list and
%   copy all the contents into your LaTeX file here.
%
% 4. Comment out the old \bibliographystyle and \bibliography commands.
%
% 5. Run LaTeX on your new file before submitting.
%
% AGU does not want a .bib or a .bbl file. Please copy in the contents of your .bbl file here.

\bibliographystyle{agufull08}
\bibliography{refs}

\newpage

%\begin{figure*}[t]  
%\noindent\includegraphics[width=40pc,angle=0]{figure3.eps}\\ 
%\caption{2010 event.  Time series of the area averaged temperature at 850 hPa and specific humidity at 850 hPa (top), relative humidity at 850
%hPa and precipitable water (center), and ERA-Interim precipitation and TRMM precipitation (bottom). The gray bars indicate the precipitation
%events. }
%  \label{timeSeries2010}
%\end{figure*}
%
%
%\begin{figure*}[t]  
%\noindent\includegraphics[width=40pc,angle=0]{figure4.eps}\\ 
%\caption{2014 event.  Time series of the area averaged temperature at 850 hPa and specific humidity at 850 hPa (top), relative humidity at 850
%hPa and precipitable water (center), and ERA-Interim precipitation and TRMM precipitation (bottom). The gray bar indicates the precipitation
%event. }
%  \label{timeSeries2014}
%\end{figure*}

%\begin{figure*}[t]  
%\noindent\includegraphics[width=40pc,angle=0]{figure8.eps}\\ 
%\caption{Sequence of maps showing the evolution of one day average precipitable water anomalies from 08-31-14 to 09-05-14.}
%  \label{PW2014}
%\end{figure*}

% See below for how to make sideways figures or tables.

\end{document}